\pdfoutput=1
\documentclass[%
 reprint,
 amsmath,amssymb,
 aps,
]{revtex4-2}
\usepackage[english]{babel}
\usepackage{braket}
\usepackage{graphicx}
\usepackage{dcolumn}
\usepackage{bm}
\usepackage{color}

\begin{document}

\preprint{APS/123-QED}

\title{Spatial entanglement engineering by pump shaping}

\author{Pauline Boucher$^1$, Hugo Defienne$^2$, and Sylvain Gigan$^{1,*}$}

\address{$^1$Laboratoire Kastler Brossel, ENS-Université PSL, CNRS, Sorbonne Université, College de France, 24 Rue Lhomond, F-75005 Paris, France\\
$^2$School of Physics and Astronomy, University of Glasgow, Glasgow G12 8QQ, UK}

\email{Corresponding author: sylvain.gigan@lkb.ens.fr}

\date{\today}

\begin{abstract}
The ability to engineer the properties of quantum optical states is essential for quantum information processing applications. Here, we demonstrate tunable control of spatial correlations between photon pairs produced by spontaneous parametric down-conversion. By shaping the spatial pump beam profile in a type-I collinear configuration, we tailor the spatial structure of coincidences between photon pairs entangled in high dimensions, without effect on intensity. The results highlight fundamental aspects of spatial coherence and hold potential for the development of quantum technologies based on high-dimensional spatial entanglement.
\end{abstract}

\maketitle


\section{Introduction}

High-dimensional quantum entanglement is an essential resource for advancing fundamental research and quantum technologies~\cite{erhard_advances_2020}. In this respect, two-photon states entangled in transverse spatial position and momentum exhibit high-dimensional entanglement and have been intensively investigated in the last decade. They are at the basis of many quantum imaging approaches, including ghost imaging~\cite{pittman_optical_1995}, sub-shot-noise~\cite{brida_experimental_2010}, resolution-enhanced~\cite{reichert_massively_2018,unternahrer_super-resolution_2018,toninelli_resolution-enhanced_2019} imaging and quantum lithography~\cite{boto_quantum_2000}. In quantum communications, high-dimensional spatial entanglement has been exploited to develop quantum cryptography protocols with higher information capacity~\cite{mirhosseini_high-dimensional_2015,erhard_twisted_2018} and increased noise resilience~\cite{ecker_overcoming_2019} by projecting photons onto spatial modes carrying angular momentum (OAM), but also by measuring them in their position-momentum bases~\cite{walborn_quantum_2006,etcheverry_quantum_2013}. All these applications strongly rely on the two-photon states properties, including their spatial entanglement structure that generally determines the capacities of the quantum-based technique. For example, it defines the information bound in certain high-dimensional quantum communication schemes~\cite{dixon_quantum_2012} and the spatial resolution in quantum imaging scheme~\cite{walborn_spatial_2010}. However, most experimental processes used to produce entangled pairs are not flexible and adjusting pairs properties to their specific use is often a challenging task. 

The most used technique for producing entangled photon pairs is spontaneous parametric down-conversion (SPDC). In SPDC, the properties of down-converted light are entirely set by the type and geometry of the non-linear crystal and the pump beam characteristics~\cite{couteau2018spontaneous}. Therefore, photon pairs with the desired joint probability distribution (JPD) may not be collected directly at the output of the crystal. The question that arises is how to manipulate independently different aspects of the JPD of entangled photon pairs produced by SPDC and, importantly, the sought-after methods that work for any pump wavelength and any nonlinear crystal. 

Numerous methods have been developed to control the type and structure of correlations of down-converted photons. The majority of them concern the time-frequency aspects of the JPD, including some based on an appropriate selection of the nonlinear crystal length and its dispersive properties~\cite{grice_eliminating_2001,kuzucu2005two,graffitti2020direct}, and others on spectral control of the pump~\cite{valencia_shaping_2007} and down-converted light~\cite{peer_temporal_2005}. To shape the spatial structure of the JPD, most approaches act directly on the down-converted photons, such as wavefront shaping~\cite{defienne_two-photon_2016,defienne_adaptive_2018-3}, quantum interferometers~\cite{zhang_engineering_2016,ferreri_spatial_2020}, metasurfaces~\cite{wang_quantum_2018,stav_quantum_2018} and rotating diffusers~\cite{lib_thermal_2020}. Recently, partial control of the spatial JPD has been achieved through spatial shaping of the pump beam profile, for example to produce entangled Airy photons~\cite{lib_spatially_2020}, compensate for optical aberrations~\cite{lib_real-time_2020-1}, and to influence its OAM~\cite{kovlakov_quantum_2018} and position-momentum degree of entanglement~\cite{defienne_spatially_2019,zhang_influence_2019}. However, a generic method to deterministically control the spatial JPD of entangled photon pairs remains a challenge. 

In this work, we propose a novel experimental approach based on wavefront shaping of the SPDC pump beam to produce entangled photon pairs with tunable spatial correlations in high dimension. 

\section{Pump-shaping and the two-photon state}
\label{sec:examples}

We place ourselves in a usual context for SPDC, that is we assume that the pump laser and down-converted fields are monochromatic, have a well defined polarization and can be faithfully described in the paraxial approximation. In this regime, the two-photon state can be expressed as \cite{Walborn_2012}:
\begin{equation}
    \ket{\Psi} = \int\int d\mathbf{q}_1 d\mathbf{q}_2 \Phi(\mathbf{q}_1, \mathbf{q}_2)\ket{\mathbf{q}_1} \ket{\mathbf{q}_2}
\end{equation}
with the normalized angular spectrum of the two-photon state $\Phi\left(\mathbf{q}_1, \mathbf{q}_2\right)$ given by
\begin{equation}
\Phi\left(\mathbf{q}_1, \mathbf{q}_2\right)=\mathcal{V}_p\left(\mathbf{q}_1 + \mathbf{q}_2\right) \mathcal{V}_c\left(\mathbf{q}_1 - \mathbf{q}_2 \right)
\label{angularspectrum}
\end{equation}
with
\begin{itemize}
\item $\mathcal{V}_p$ the normalized angular spectrum of the pump
\item $\mathcal{V}_c \left(\mathbf{q}\right)= \frac{1}{\pi}\sqrt{\frac{2L}{K}} sinc\left( \frac{L\left|\mathbf{q}\right|^2}{4K}\right)$ with $L$ the length of the nonlinear crystal and $K$ wave number of the pump field
\end{itemize}
It is straightforward to see in equation \ref{angularspectrum} that the angular spectrum of the pump directly shapes that of the two-photon state. In this work, we aim at manipulating the pump angular spectrum through spatial pump shaping in order to engineer the angular spectrum of the two-photon state. For this shaping to be observed, it should modulate frequencies such that $\mathbf{q}_1 - \mathbf{q}_2$ does not belong to the kernel of $\mathcal{V}_c$. Depending on the crystal parameters, this condition offers some freedom for the observation of spatial modulations. As an illustration, we now present an analytical solution for the case of a Bessel-Gauss pump beam \cite{schneeloch_position-momentum_2016, doi:10.1080/0010751042000275259}.

First, in many SPDC experimental conditions, the $sinc$ function in \ref{angularspectrum} can be approximated by a Gaussian function~\cite{fedorov_gaussian_2009}, such that
\begin{equation}
\mathcal{V}_c \left(\mathbf{q}\right) \simeq \delta \exp\left(-\frac{\delta^2\left|\mathbf{q}\right|^2}{2}\right)
\end{equation}
with $\delta \simeq 0.257 \sqrt{L/4K}$. At its waist, i.e. $z=0$, a Bessel-Gauss beam of order $l$ can be expressed as 
\begin{equation}
BG(r, \phi,0;l) = A J_l(k_r r)\exp(il\phi) \exp(-\frac{r^2}{w_g^2}).
\end{equation}
with $J_l$ the $l^{\text{th}}$ Bessel function of the first kind. Its angular spectrum is given by \cite{Vaity:15}: 
\begin{align}
\mathcal{V}^{BG}_{p}(q, \phi_q, 0;l) =&i^{l-1} \frac{w_g}{w_0}\exp(il\phi_q)\exp(-\frac{q^2 + k_r^2}{w_0^2})\nonumber\\
& \times I_l\left(\frac{2k_r q}{w_0^2}\right)
\end{align}
with $w_0 = 2/w_g$ and $I_l$ is the $l^{\text{th}}$ order modified Bessel function of the first kind. Thus, for a pump beam of the $0^{\text{th}}$ order Bessel-Gauss beam, the two-photon state can be expressed analytically:
\begin{align}
\Phi\left(\mathbf{q}_1, \mathbf{q}_2\right)&=\frac{\pi \delta}{k_r}\frac{-i w_g^2}{2} \exp(-4\frac{\left|\mathbf{q}_1 + \mathbf{q}_2\right|^2 + k_r^2}{w_g^2}) \nonumber\\
& \times I_0\left(\frac{8k_r \left|\mathbf{q}_1 + \mathbf{q}_2\right|}{w_g^2}\right)  \exp\left(-\frac{\delta^2\left|\mathbf{q}_1 - \mathbf{q}_2\right|^2}{2}\right).
\end{align}
Even though not all angular spectrum functions allow for such analytical decompositions, numerical integration easily allows the tailoring of the pump spatial profile for given two-photon state spatial properties. However, the space of accessible two-photons states through pump shaping is fundamentally limited in terms of symmetries by the fact that the angular spectrum is a function of $\mathbf{q}_1 +  \mathbf{q}_2$.

\section{Experiment}
\begin{figure}[htbp]
\centering
\includegraphics[width=\linewidth]{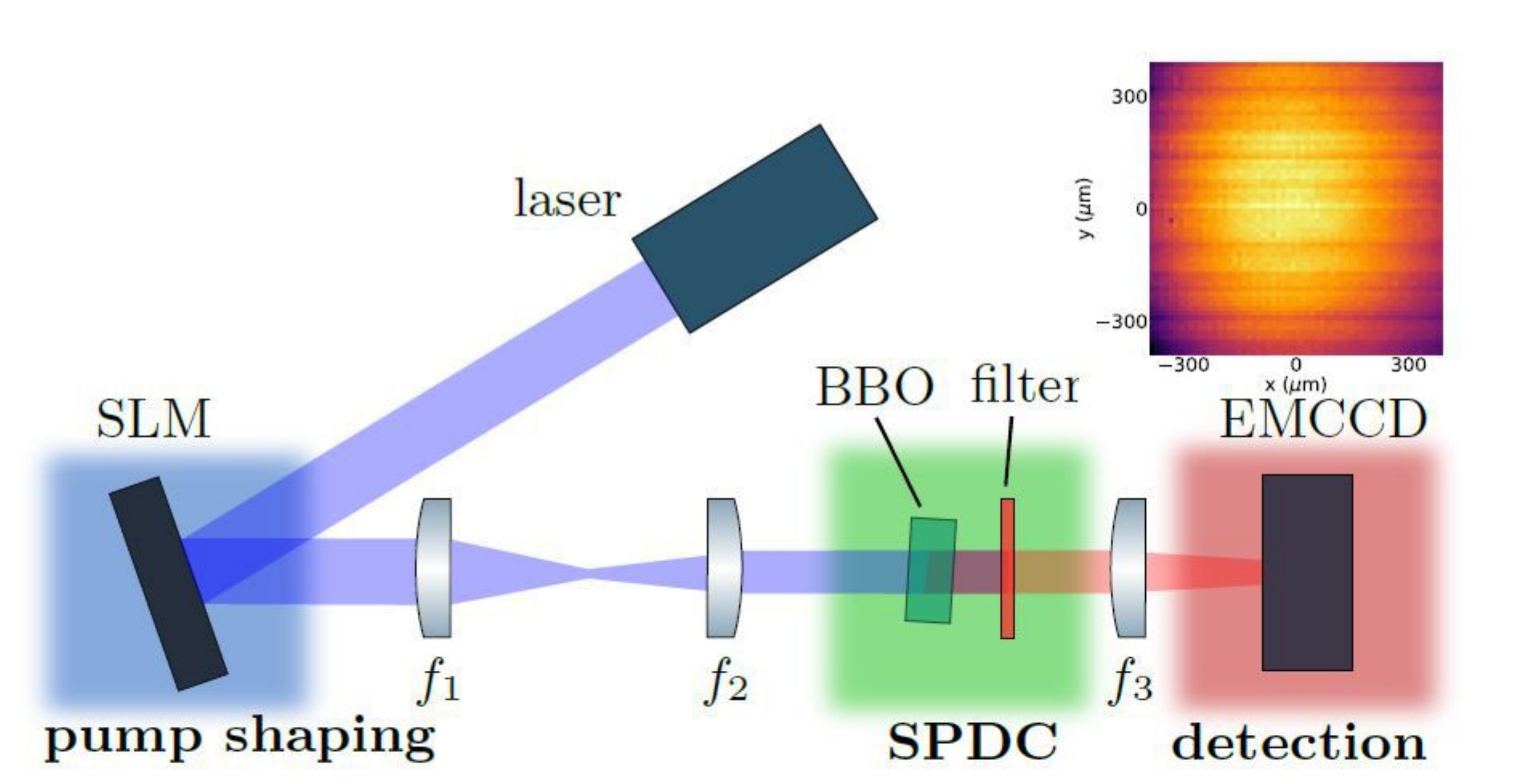}
\caption{A spatially monomode, low-power, continuous wave laser whose angular spectrum is tailored using a SLM and a 4-f system and injected into a BBO crystal in a type-I SPDC configuration. Using two different lenses ($f_3^a$ and $f_3^b$), the near- and far-field of the crystal can be measured.}
\label{fig:schema-exp}
\end{figure}
To observe the effect of pump shaping on the two-photon state, we used the experimental setup described in Figure~\ref{fig:schema-exp}. A pump beam at 405 nm is spatially modulated using a spatial light modulator (SLM) and imaged onto the front surface of a $\beta$-barium borate (BBO) crystal. The pump beam is filtered out using a narrow band filter at 810nm. The distance between the crystal and the camera sensor $d$ was finely tuned so that by positioning the imaging lens of focal length $f_3$ precisely half-way and switching between two different lenses $a$ and $b$, we could image the near-field of the crystal plane (2f-2f configuration with $f_3^a = d/4$) or its far-field (f-f configuration with $f_3^b= d/2$). We used an EMCCD camera (Andor iXon) operated in photon counting mode to measure photon correlations, following the procedure described in \cite{defienne_general_2018-2}.

\section{Results}
The spatial profile of the pump beam was engineered to produce a variety of two-photon states.
\begin{figure}[htbp]
\centering
\includegraphics[width=\linewidth]{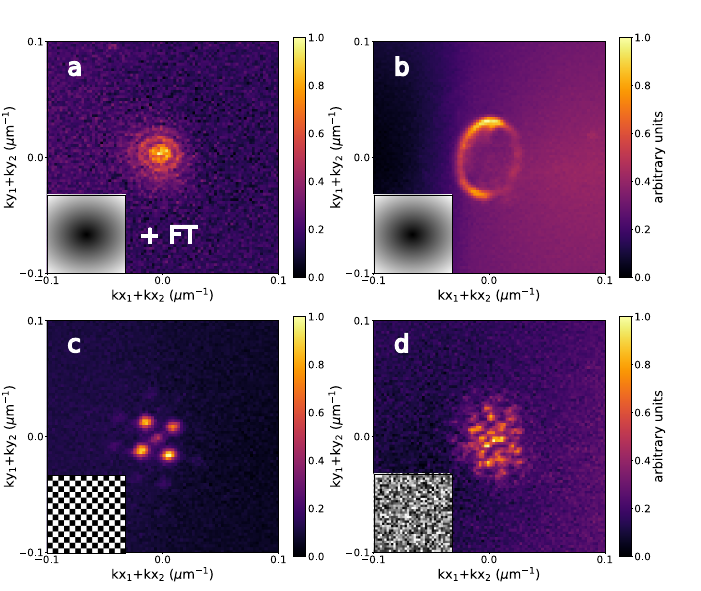}
\caption{Autoconvolutions for different pump profiles with SLM profiles in inset. In (a), the pump beam is shaped to the Fourier transform of a Bessel beam, i.e. ring-shaped and, in (b), directly as a Bessel beam. In (c), a checker board is displayed, while in (d) we display a random pattern. In each case, the autoconvolution measured for the down-converted photons corresponds to the pump beam's angular spectrum.}
\label{fig:fig-a}
\end{figure}
By displaying an axicon profile on the SLM, the beam pumping the crystal can be shaped into a Bessel-Gauss beam \cite{PhysRevLett.58.1499, GORI1987491}. In Figure~\ref{fig:fig-a} (a), we illuminate the crystal with a Fourier-transformed Bessel-Gauss beam and observe that the down-converted photons' autoconvolution is shaped into a Bessel function. The autoconvolution is defined as the projection of the joint probability distribution of photon pairs in the sum-coordinate basis $\{ x_1+x_2,y_1+y_2\}$. Likewise, when pumping directly with a Bessel-Gauss beam (Figure~\ref{fig:fig-a}(b)), the obtained autoconvolution is the Fourier transform of a Bessel function, which is a ring-shaped function. By displaying a checker-board pattern (Figure~\ref{fig:fig-a}(c)) or a random pattern (Figure~\ref{fig:fig-a}(d)), the autoconvolution is likewise shaped as the Fourier transform of the pumping angular spectrum.

\begin{figure}[htbp]
\centering
\includegraphics[width=\linewidth]{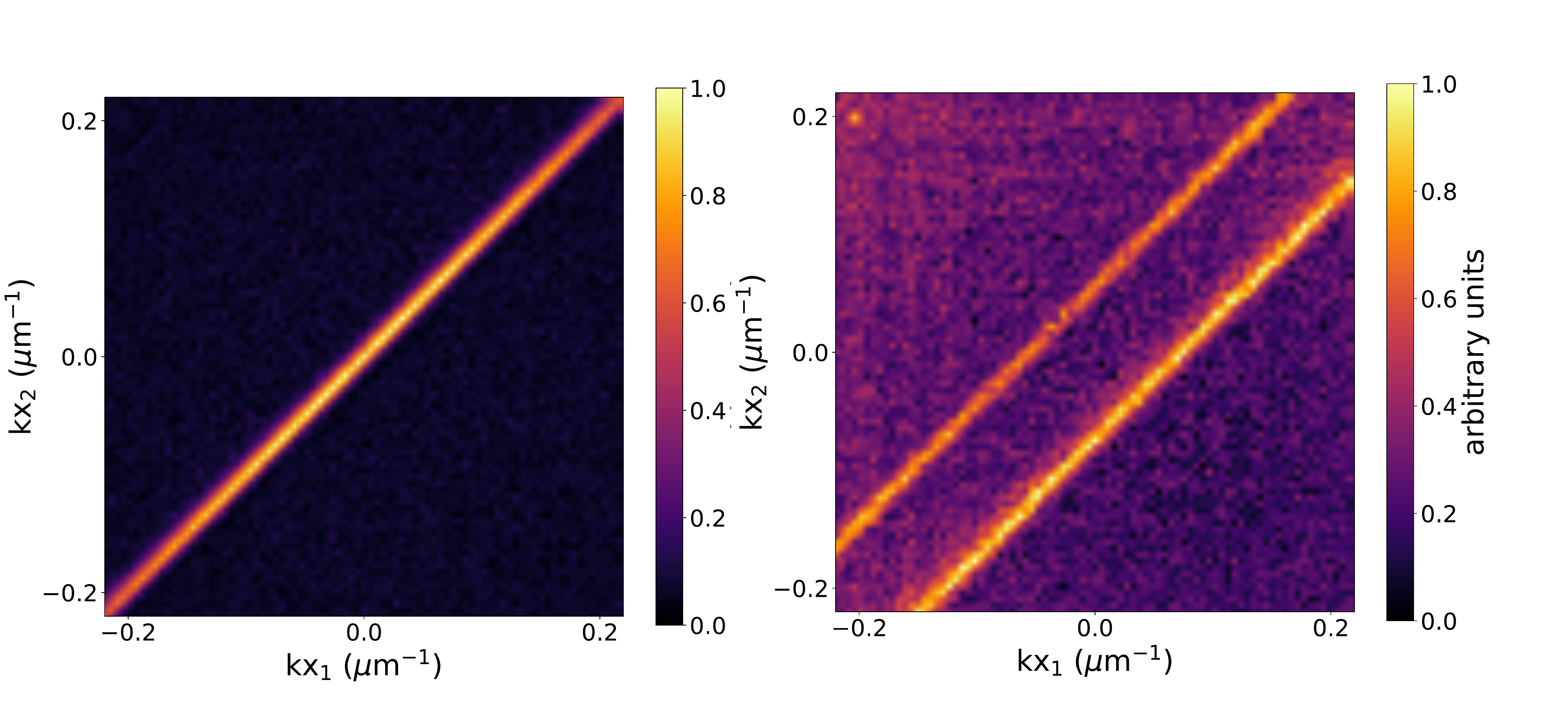}
\caption{Average of the photon correlations between symmetric rows of the cameras for all $k_{y1} = -k_{y1}$. (a): Gaussian pump. Clear anticorrelation in the momentum of the down-converted photons is present as expected. (b): Bessel-Gauss pump. The antidiagonal is here split in two, demonstrating the possibility of correlation properties engineering.}
\label{fig:fig-b}
\end{figure}

To further illustrate correlations engineering of the down-converted photons, we also measured the photon correlations between pairs of symmetric rows of the camera, as shown in Figure~\ref{fig:fig-b}. An element ($kx_1$, $kx_2$) therefore corresponds to the joint probability of detecting one photon at $\mathbf{k}_{1} = (k_{x1}, k_{y1})$ together with the second photon at $\mathbf{k}_2 = (k_{x2} , -k_{y1})$ (summed over all $k_{y1}$). In Figure~\ref{fig:fig-b} (a), the strong antidiagonal is a signature of momentum conservation between photons produced by SPDC in the classical case of a Gaussian pump. In the case of a Bessel-Gauss pump (Figure~\ref{fig:fig-b}(b)), we clearly observe shaping of the photon correlations by a split of the antidiagonal. Indeed, in a Bessel beam, pump photons all possess the same momentum in absolute value: having fixed $k_{y1}+k_{y2}=0$, the conservation of momentum allows for two different solutions $k_{x2} = - k_{x1} \pm |k_{p}|$, which correspond to the two antidiagonals.
\section{Conclusion}
We demonstrate a versatile approach for spatial correlations engineering of SPDC photon pairs. It is fully compatible with, and could easily be integrated in any conventional type-I SPDC setup, and should also readily be applicable to other pair production processes such as type-II SPDC and atomic vapor systems. This approach also exhibits the advantage of maintaining the down-converted photon rate: indeed, the losses introduced are only affecting the pump beam and can easily be compensated by an increase in pump power. Tomography of the down-converted field, in the fashion of \cite{roslund_wavelength-multiplexed_2014}, is necessary to completely characterize the produced states.

\textbf{Funding:}
This work was funded by the European Research Council (ERC; H2020, SMARTIES-724473). SG acknowledges support from the Institut Universitaire de France. H.D. acknowledges funding from the European Union’s Horizon 2020 research and innovation program under the Marie Skłodowska-Curie grant no. 840958.


\bibliography{biblio}

\end{document}